\documentclass[aps, prl, showpacs, twocolumn, superscriptaddress]{revtex4}

\usepackage{graphicx}
\usepackage{dcolumn}
\newcommand{\dash}{\multicolumn{1}{c}{---}}

\begin{document}

\title{3D Collapse of Rotating Stellar Iron Cores in General Relativity including 
Deleptonization and a Nuclear Equation of State}

\author{C.~D.~Ott}
\affiliation{Max-Planck-Institut f\"ur Gravitationsphysik,
  Albert-Einstein-Institut, Am M\"uhlenberg~1, D-14476 Potsdam,
  Germany}

\author{H.~Dimmelmeier}
\affiliation{Max-Planck-Institut f\"ur Astrophysik,
  Karl-Schwarzschild-Strasse~1, D-85741 Garching, Germany}

\author{A.~Marek}
\affiliation{Max-Planck-Institut f\"ur Astrophysik,
  Karl-Schwarzschild-Strasse~1, D-85741 Garching, Germany}

\author{H.-T.~Janka}
\affiliation{Max-Planck-Institut f\"ur Astrophysik,
  Karl-Schwarzschild-Strasse~1, D-85741 Garching, Germany}

\author{I.~Hawke}
\affiliation{School of Mathematics, University of Southampton,
  Southampton SO17 1BJ, UK}

\author{B.~Zink}
\affiliation{Max-Planck-Institut f\"ur Astrophysik,
  Karl-Schwarzschild-Strasse~1, D-85741 Garching, Germany}
\affiliation{Center for Computation and Technology, Louisiana State
  University, Baton Rouge, LA 70803, USA}

\author{E.~Schnetter}
\affiliation{Center for Computation and Technology, Louisiana State
  University, Baton Rouge, LA 70803, USA}

%%%%%%%%%%%%%%%%%%%%%%%%%%%%%%%%%%%%%%%%%%%%%%%%%%%%%%%%%%%%%%%%%%%%%%%%%%%%%%%%%%%
%%%%%%%%%%%%%%%%%%%%%%%%%%%%%%%%%%%%%%%%%%%%%%%%%%%%%%%%%%%%%%%%%%%%%%%%%%%%%%%%%%%
%%%%%%%%%%%%%%%%%%%%%%%%%%%%%%%%%%%%%%%%%%%%%%%%%%%%%%%%%%%%%%%%%%%%%%%%%%%%%%%%%%%

\begin{abstract}
  We present results from the first 2D and 3D simulations of the
  collapse of rotating stellar iron cores in general relativity
  employing a finite-temperature equation of state and an approximate
  treatment of deleptonization during collapse.  We compare fully
  nonlinear and conformally flat spacetime evolution methods and find
  that the conformally flat treatment is sufficiently accurate for the
  core-collapse supernova problem. We focus on the gravitational wave
  (GW) emission from rotating collapse, core bounce, and early
  postbounce phases. Our results indicate that the GW signature of
  these phases is much more generic than previously estimated. In
  addition, we track the growth of a nonaxisymmetric instability of
  dominant $ m = 1 $ character in one of our models that leads to
  prolonged narrow-band GW emission at $ \sim 930 \mathrm{\ Hz} $ over
  several tens of milliseconds.
\end{abstract}

\maketitle

%%%%%%%%%%%%%%%%%%%%%%%%%%%%%%%%%%%%%%%%%%%%%%%%%%%%%%%%%%%%%%%%%%%%%%%%%%%%%%%%%%%
%%%%%%%%%%%%%%%%%%%%%%%%%%%%%%%%%%%%%%%%%%%%%%%%%%%%%%%%%%%%%%%%%%%%%%%%%%%%%%%%%%%
%%%%%%%%%%%%%%%%%%%%%%%%%%%%%%%%%%%%%%%%%%%%%%%%%%%%%%%%%%%%%%%%%%%%%%%%%%%%%%%%%%%

\emph{Introduction.}---%
For more than two decades astrophysicists have struggled to compute
the gravitational wave (GW) signal produced by rotating stellar core
collapse and the subsequent supernova evolution. Besides the
coalescence of black hole and neutron star binaries, core-collapse
events are considered to be among the most promising sources of
detectable GWs. Theoretical predictions are still hampered by three
major problems: (i) The rotational configuration prior to
gravitational collapse is still uncertain since multi-D evolutionary
calculations of rotating massive stars have not yet been performed;
(ii) reliable waveform estimates require a general relativistic (GR)
treatment, since both high densities and velocities in combination
with strong gravitational fields are encountered in this problem; and
(iii) an adequate treatment of the nuclear equation of state (EOS) and
the neutrino microphysics/radiative transfer is crucial for obtaining
realistic collapse, bounce, and postbounce dynamics and waveforms. GW
emission from core-collapse supernovae may arise from rotating
collapse and bounce, postbounce neutrino-driven convection,
anisotropic neutrino emission, nonaxisymmetric rotational
instabilities of the protoneutron star (PNS), or from the recently
proposed PNS core g-mode oscillations. Previous estimates of the GW
signature of core-collapse supernovae have either relied on Newtonian
simulations~\cite{moenchmeyer_91_a,zwerger_97_a,rampp_98_a,kotake_04_a,ott_04_a,ott_06_a}
(to some extent approximating GR effects~\cite{mueller_04_a})
or GR simulations with simplified analytic (so-called hybrid) EOSs and
no neutrino treatment~\cite{dimmelmeier_02_a_b,shibata_04_a,shibata_05_a,cerda_05_a}.
Depending on the rotation strength, softness of the EOS at subnuclear
densities, and inclusion of GR effects, the collapse dynamics and
accordingly the GW signature can differ significantly.

Here we present new results from GR simulations, focussing on the
rotating collapse, bounce, and early postbounce phases. These are the
\emph{first-ever} multi-D simulations in GR with presupernova models
from stellar evolution calculations, a finite-temperature nuclear EOS,
and a simple, but effective treatment of electron capture and neutrino
radiation effects during collapse. In this way we obtain the most
accurate estimates of the GW signature of rotating stellar core
collapse in full GR to date.

%%%%%%%%%%%%%%%%%%%%%%%%%%%%%%%%%%%%%%%%%%%%%%%%%%%%%%%%%%%%%%%%%%%%%%%%%%%%%%%%%%%
%%%%%%%%%%%%%%%%%%%%%%%%%%%%%%%%%%%%%%%%%%%%%%%%%%%%%%%%%%%%%%%%%%%%%%%%%%%%%%%%%%%
%%%%%%%%%%%%%%%%%%%%%%%%%%%%%%%%%%%%%%%%%%%%%%%%%%%%%%%%%%%%%%%%%%%%%%%%%%%%%%%%%%%

\emph{Method and Initial Models.}---%
We perform all 3D simulations in full 3\,+\,1 GR using the
\textsc{Cactus} infrastructure~\cite{cactus_code}, Cartesian coordinates,
and mesh refinement provided by the \textsc{Carpet}
driver~\cite{schnetter_04_a}. The only assumption on symmetry is
reflection invariance with respect to the equatorial plane. Spacetime
is evolved using the BSSN formulation~(see, e.g., \cite{baumgarte_03_a})
and we fix the gauge freedom by $ 1 + \log $ slicing and by a
hyperbolic shift\cite{shibata_03_a}. We use the hydrodynamics code
\textsc{Whisky}~\cite{baiotti_05_a}, which implements the equations of
GR hydrodynamics via finite-volume methods. Typical simulation grids
extend to $ 3000 \mathrm{\ km} $ and use 9 refinement levels. The
central resolution is $ \sim 350 \mathrm{\ m} $. In addition, we
perform axisymmetric (2D) simulations for all models using the
\textsc{CoCoNuT} code~\cite{dimmelmeier_02_a_b,dimmelmeier_05_a}, which
approximates GR by the conformal flatness condition
(CFC). \textsc{CoCoNuT} utilizes spherical coordinates with 250
logarithmically spaced radial and 45 equidistant angular zones,
covering $ 90^\circ $, and a central resolution of
$ \sim 250 \mathrm{\ m} $. Resolution test calculations with
both CoCoNuT and CCW do not yield significant qualitative or
quantitative differences. We extract GWs using
a variant of the Newtonian quadrupole formula~\cite{shibata_04_a}.

We employ the finite-temperature nuclear EOS of Shen et
al.~\cite{shen_98_a}. Deleptonization is implemented
as proposed by \cite{liebendoerfer_05_a}: 
The electron fraction
$ Y_e $ is parameterized as a function of density 
based on data from spherically symmetric radiation-hydrodynamics
calculations with standard electron capture rates~\cite{buras_06}.
After core bounce, $ Y_e $ is passively advected and further lepton
loss is neglected, but neutrino pressure contributions continue to be
taken into account above trapping density~\cite{liebendoerfer_05_a}.

In this study we focus on the collapse of massive presupernova iron
cores with at most moderate differential rotation, and rotation rates
that may be too fast to match garden-variety pulsar birth spin
estimates~\cite{heger_05_a,ott_06_b}, but could be relevant in the
collapsar-type gamma-ray burst context. As
initial data we use the non-rotating $ 20\,M_\odot $ presupernova
model s20 of Woosley et al.~\cite{woosley_02_a} which we force to
rotate according to the rotation law discussed
in~\cite{ott_04_a,dimmelmeier_02_a_b}. We parameterize our models in
terms of the differential rotation parameter $ A $ and the initial
ratio of rotational kinetic to gravitational binding energy
$ \beta_\mathrm{i} = T / |W| $.  In addition, we perform a calculation
with the $ 20\,M_\odot $ model E20A of Heger et al.~\cite{heger_00_a},
which was evolved to the onset of collapse with a 1D treatment of
rotation. In Table~\ref{tab:model_summary} we summarize the model
parameters.

\begin{table}
  \caption{Summary. $ \rho_\mathrm{b} $ is the density at
    bounce, the maximum characteristic GW strain
    $ h_\mathrm{char,max} $ is at a distance of $ 10 \mathrm{\ kpc} $,
    and $ E_\mathrm{gw} $ is the energy emitted in GWs. Values for
    $ \mathrm{E20A}_\mathrm{pb} $ include GW emission from
    late-time 3D dynamics.}
  \begin{ruledtabular}
    \begin{tabular}{ldddddd}
      Model
      & \multicolumn{1}{c}{$ A $}
      & \multicolumn{1}{c}{$ \beta_\mathrm{i} $}
      & \multicolumn{1}{c}{$ \beta_\mathrm{b} $}
      & \multicolumn{1}{c}{$ \rho_\mathrm{b} $}
      & \multicolumn{1}{c}{$ h_\mathrm{char,max} $}
      & \multicolumn{1}{c}{$ E_\mathrm{gw} $} \\
      & \multicolumn{1}{c}{\scriptsize[$ 10^8 \mathrm{\,cm} $]}
      & \multicolumn{1}{c}{\scriptsize[\%]}
      & \multicolumn{1}{c}{\scriptsize[\%]}
      & \multicolumn{1}{c}{\scriptsize$ \displaystyle \left[\!
        \frac{10^{14} \mathrm{\,g}}{\mathrm{cm}^3} \!\right] $}
      & \multicolumn{1}{c}{\scriptsize[$ 10^{-21} $]}
      & \multicolumn{1}{c}{\scriptsize[$ 10^{-9} M_\odot c^2 $]} \\ [0.7 em]
      \hline
      $\rule{0 em}{1 em}$%
      s20A1B1 & 50.0  & 0.25 &  0.90 & 3.29 &  1.46 &  0.6 \\
      s20A1B5 & 50.0  & 4.00 & 10.52 & 2.90 &  9.68 & 26.9 \\
      s20A2B2 &  1.0  & 0.50 &  6.72 & 3.07 &  8.77 & 22.0 \\
      s20A2B4 &  1.0  & 1.80 & 16.33 & 2.35 &  4.28 &  9.4 \\
      s20A3B3 &  0.5  & 0.90 & 16.57 & 2.33 &  4.58 & 12.4 \\
      E20A    & \dash & 0.37 & 11.31 & 2.79 & 12.18 & 36.9 \\
      $ \mathrm{E20A}_\mathrm{pb} $ & & & & & 24.23 & 75.4
    \end{tabular}
  \end{ruledtabular}
  \label{tab:model_summary}
\end{table}

%%%%%%%%%%%%%%%%%%%%%%%%%%%%%%%%%%%%%%%%%%%%%%%%%%%%%%%%%%%%%%%%%%%%%%%%%%%%%%%%%%%
%%%%%%%%%%%%%%%%%%%%%%%%%%%%%%%%%%%%%%%%%%%%%%%%%%%%%%%%%%%%%%%%%%%%%%%%%%%%%%%%%%%
%%%%%%%%%%%%%%%%%%%%%%%%%%%%%%%%%%%%%%%%%%%%%%%%%%%%%%%%%%%%%%%%%%%%%%%%%%%%%%%%%%%

\begin{figure}
  \includegraphics[width=8.7cm]{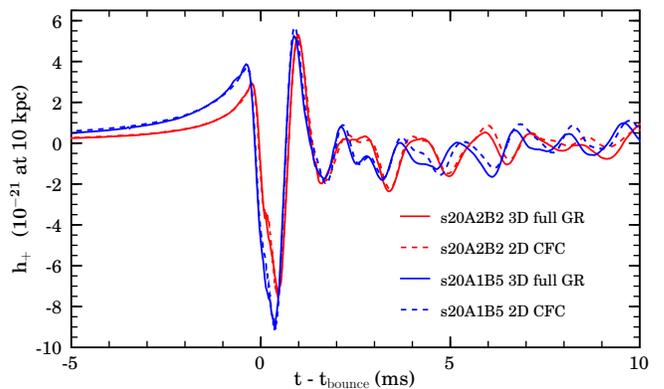}
  \caption{GW strain $ h_+ $ along the equator for models s20A2B2 and
    s20A1B5. We compare 2D-CFC and 3D-full-GR results.}
  \label{fig:cfc_cactus}
\end{figure}

\emph{Results.}---%
In Fig.~\ref{fig:cfc_cactus} we compare GW signals computed with
\textsc{CoCoNuT} in 2D-CFC and those computed with our 3D-full-GR
approach. Model s20A2B2 (red lines) is a moderate rotator with a
$ \beta_\mathrm{i} = 0.50\% $, rotating rigidly in its central
region. It stays axisymmetric throughout its numerical evolution. The
agreement of 2D-CFC with 3D-full-GR is excellent for this model: Both
waveforms match almost perfectly at bounce and during the very early
postbounce phase. A few ms after bounce, when convection in the region
behind the stalling shock sets in due to a negative entropy gradient,
the signals begin to differ quantitatively while remaining in
phase. We attribute this small mismatch to the choice of coordinate
grids and to differences in the growth and scale of vortical
postbounce motions between 2D and 3D. Model s20A1B5 rotates with
constant $ \Omega $ in the entire core. Despite its very large
$ \beta_\mathrm{i} = 4\% $ it remains essentially axisymmetric during
the time covered by our simulation, since most of its angular momentum
is attached to material at large radii that falls inward and spins up
only slowly. The waveforms in CFC and full GR agree very well. Again,
both waveforms match best for the strong burst related to core bounce
during which more than $ \sim 90\% $ of the total GW energy are
emitted in an axisymmetric model. The overall excellent agreement of
CFC with full GR confirms results of~\cite{shibata_04_a,cerda_05_a} and
proves that CFC is a very good approximation to full GR in the
core-collapse scenario.

\begin{figure}
  \includegraphics[width=8.7cm]{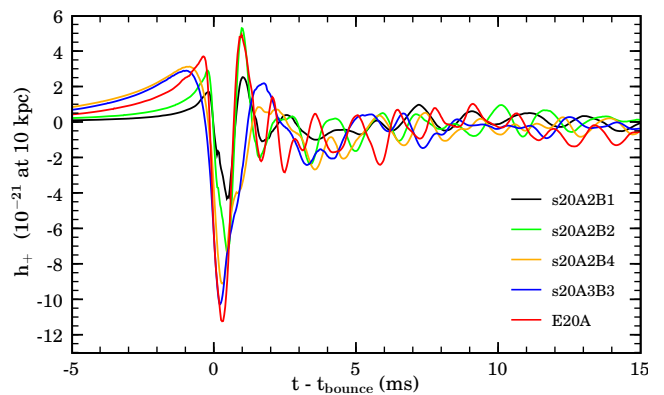}
  \caption{Equatorial GW strain $ h_+ $ for representative models.}
  \label{fig:waveforms}
\end{figure}

In Fig.~\ref{fig:waveforms} we present waveforms of models with
varying initial degree of differential rotation $ A $ and rotation
rate $ \beta_\mathrm{i} $. We find that the inclusion of a
microphysical finite-temperature EOS and of electron capture during
collapse yields results that differ considerably from those obtained
in previous, less sophisticated studies. Fig.~\ref{fig:waveforms}
exemplifies that largely independent of the initial rotational
configuration, the GW signal of rotating collapse has 
a \emph{generic shape}: a slow rise in the pre-bounce phase, a large 
negative amplitude at bounce when the motion of
the inner core is reversed, followed by
a ring-down. This so-called ``Type~I'' signature corresponds to a
baryonic pressure dominated
bounce~\cite{moenchmeyer_91_a,zwerger_97_a,dimmelmeier_02_a_b,ott_04_a}.
Thus all our models undergo core bounce dominated by the stiffening of
the EOS at nuclear density.

This is in stark contrast to the studies using the hybrid 
EOS~\cite{zwerger_97_a,dimmelmeier_02_a_b,shibata_04_a,shibata_05_a},
where initial models with rotation rates in the range investigated
here develop sufficient centrifugal support during contraction to stop
collapse at subnuclear densities, resulting in several consecutive
centrifugal bounces separated by phases of coherent re-expansion of
the inner core. While in GR models exhibiting such multiple
centrifugal bounce and the corresponding ``Type~II'' GW signals are
only rarer compared to Newtonian gravity~\cite{dimmelmeier_02_a_b}, we
do not observe \emph{any} such model in this study.  An evident
example is model s20A2B4: In previous studies without detailed
microphysics, the corresponding model with identical initial rotation
parameters (A2B4G1) showed clear ``Type~II'' behavior in both
Newtonian and GR calculations~\cite{zwerger_97_a,dimmelmeier_02_a_b}.

The suppression of the multiple centrifugal bounce scenario is due to
two physical effects: (i) In contrast to the simple hybrid EOS, in our
case the mass and dynamics of the inner core (which is most important
for the GW emission) is not merely determined by the adiabatic index
$ \gamma = d \ln P / d \ln \rho $ (at constant entropy) of the
EOS, but also by deleptonization during collapse.  This leads to
\emph{considerably smaller inner cores with less angular momentum and
  weaker pressure support}. (ii) Since multiple centrifugal bounce was
observed for a model with initially moderately fast rotation in a
previous Newtonian study~\cite{moenchmeyer_91_a} where both a
microphysical finite-temperature EOS and a deleptonization scheme were
employed, the absence of this collapse type in our study is not only
caused by microphysical effects, but also by the effectively
\emph{stronger} gravity in GR. This is in accordance with simulations
using the simple hybrid EOS~\cite{dimmelmeier_02_a_b}. A detailed
analysis of the interplay and quantitative influence of the above two
effects responsible for the elimination of multiple centrifugal bounce
in the rotational stellar core-collapse scenario will be presented in
a future publication.

\begin{figure}
  \includegraphics[width=8.1cm]{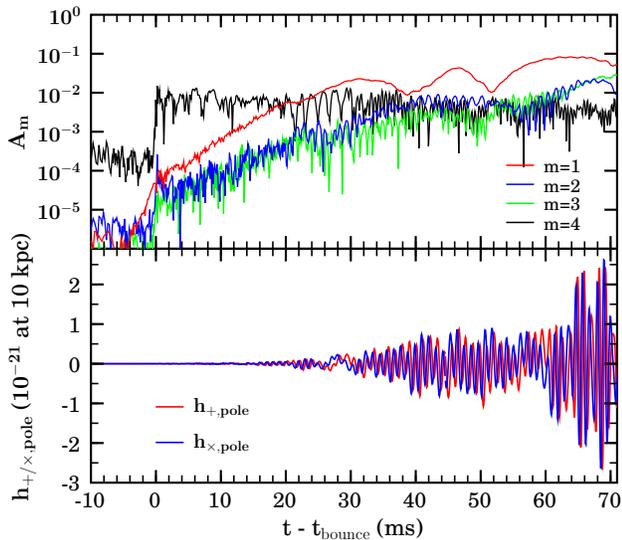}
  \caption{Normalized mode amplitudes $ A_m $ at postbounce times
    (upper panel) and GW strains $ h_+ $ and $ h_\times $ along the
    poles (lower panel) for model E20A.}
  \label{fig:bar_mode_signal}
\end{figure}

Model E20A possesses the largest GW amplitude of all our models. In
addition, it reaches a high $ \beta_\mathrm{b} $ at core bounce (see
Table~\ref{tab:model_summary}) and settles at a postbounce
$ \beta_\mathrm{f} $ of $ \sim 9\% $. In a previous Newtonian 
study, \cite{ott_05_a} have found a low-$ T / |W| $
nonaxisymmetric instability for a PNS with similar
$ \beta_\mathrm{f} $. In order to verify their findings, we trace
the evolution of model E20A to $ 70 \mathrm{\ ms} $ after bounce and
perform an analysis of azimuthal density modes $ \propto e^{im\varphi} $
in the equatorial plane by computing complex Fourier amplitudes
$ C_m = \frac{1}{2\pi} \int^{2\pi}_0 \! \rho(\varpi, \varphi, z = 0)
\, e^{im\varphi} \, d\varphi $ on rings of constant coordinate radius
with respect to the coordinate center of mass. The latter stays
within the innermost zone at all times.
In the top panel of Fig.~\ref{fig:bar_mode_signal} we display the
normalized mode amplitudes $ A_m = \frac{|C_m|}{C_0} $ extracted at
$ 15 \mathrm{\ km} $ radius. Without adding artificial seed
perturbations to model E20A, discretization errors trigger
$ m = \{1,2,3\} $ modes, which rise to a level of $ \sim 10^{-5} $
during the $ \sim 220 \mathrm{\ ms} $ collapse phase.
After bounce, the m=1 mode exhibits the fastest growth. This
growth on a dynamical time scale, lasting over tens of ms until
saturation, is closely followed by a growth of $ m = \{2,3\} $ daughter
modes~\cite{ott_05_a,ou_06_a}. Note that the $ m = 4 $ Cartesian grid
mode starts out on the $ \sim 10^{-4} $ level and remains constant
until the plunge phase of collapse during which all modes are
amplified. After core bounce, model E20A remains dynamically stable to
the $ m = 4 $ grid mode. In the lower panel of
Fig.~\ref{fig:bar_mode_signal} we plot the  GW strains $ h_+ $ and
$ h_\times $ as seen along the polar  axis. The rotational symmetry of
E20A before and early after bounce times is reflected in the fact that
$ h_+ $ and $ h_\times $ are practically zero until E20A develops
considerable nonaxisymmetry at $ \sim 25 \mathrm{\ ms} $ after bounce
when the $ m = 1 $ mode becomes dominant and its $ m = 2 $,
GW-emitting harmonic reaches a sizable amplitude. In remarkable
agreement with expectations for a spinning bar, $ h_+ $ and $ h_\times
$ oscillate at the same frequency ($ \sim 930 $ Hz) and are
phase-shifted by a quarter cycle.

%%%%%%%%%%%%%%%%%%%%%%%%%%%%%%%%%%%%%%%%%%%%%%%%%%%%%%%%%%%%%%%%%%%%%%%%%%%%%%%%%%%
%%%%%%%%%%%%%%%%%%%%%%%%%%%%%%%%%%%%%%%%%%%%%%%%%%%%%%%%%%%%%%%%%%%%%%%%%%%%%%%%%%%
%%%%%%%%%%%%%%%%%%%%%%%%%%%%%%%%%%%%%%%%%%%%%%%%%%%%%%%%%%%%%%%%%%%%%%%%%%%%%%%%%%%

\emph{Discussion.}---%
Our results indicate that the GW signature of the collapse, 
bounce, and early postbounce phases of the core-collapse supernova
evolution is much more generic than previously thought. We find that
the dynamics of core bounce are dominated by mainly gravity and
microphysics, reducing the importance of centrifugal support for the
rotation rates considered here. Importantly, for our model set we do
not observe rotationally induced multiple core bounce at subnuclear
density as proposed by previous studies that did not include a
microphysical finite-temperature EOS and electron capture treatment in
combination with GR. Thus we predict that the core-bounce waveform of
models in a large parameter space of initial rotation rate and degree
of differential rotation will likely both qualitatively and
quantitatively resemble those presented in Fig.~\ref{fig:waveforms}.

Model E20A, which we evolve to later postbounce times, exhibits the
dynamical growth of a nonaxisymmetric low-$ T / |W| $ corotation-type
$ m = 1 $ instability \cite{saijo_06_a,ou_06_a,ott_05_a}. We also find
$ m = \{2,3\} $ contributions and thus significant GW emission from
the quadrupole components of these modes. We emphasize that we observe
this instability not only in E20A, but also in other models with
comparable values of $ \beta_\mathrm{f} $. Our results, which remove
the limitations of previous 
studies~\cite{centrella_01_a,shibata_02_a,shibata_05_a,ott_05_a},
demonstrate that the development of nonaxisymmetric structures is
neither limited to Newtonian gravity, simple matter models,
equilibrium configurations, nor high values of $ \beta $,  
but may rather be a phenomenon occurring generically in differentially
rotating compact stars.

\begin{figure}
  \includegraphics[width=7.9cm]{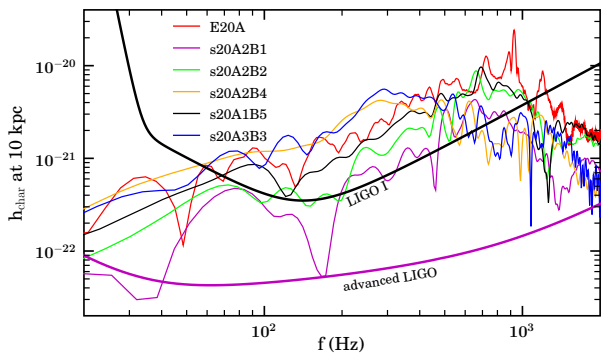}
  \caption{Spectra of the characteristic GW strain
    $ h_\mathrm{char} $ of all models and the LIGO (optimal) rms noise
    curves~\cite{shoemaker_06_a}.}
  \label{fig:spectrum}
\end{figure}

For the GW signals from the axisymmetric collapse and core-bounce
phase we obtain peak amplitudes of up to $ h \sim 10^{-20} $ at
$ 10 \mathrm{\ kpc} $, while the nonaxisymmetric structures in model
E20A developing later emit GWs with $ h $ only a factor $ \sim 5 $
smaller. However, since the latter emission process operates over
several tens of ms, the total energy $ E_\mathrm{gw} $ emitted in GWs
is \emph{larger} than that from the core-bounce signal. This is
evident in Fig.~\ref{fig:spectrum}, where we display the
characteristic GW strain spectra $ h_\mathrm{char} = R^{-1}
\sqrt{2 \pi^{-2} dE_\mathrm{gw} / df} $~\cite{ott_04_a} for all models,
evolving E20A for $ 70 \mathrm{\ ms} $ after bounce (see also
Table~\ref{tab:model_summary}). Considering only the
core-bounce waveforms, $ h_\mathrm{char} $ has its maximum between
$ 300 $ and $ 800 \mathrm{\ Hz} $, while for model E20A it peaks at
$ \sim 930 \mathrm{\ Hz} $, the frequency of the GW-emitting
component of its nonaxisymmetric structures. We conclude that the
core-bounce GW signals of all models investigated here may be
detectable by current and future LIGO-class detectors from anywhere in
the Milky Way. Models that develop nonaxisymmetric instabilities may
be detectable out to much larger distances if the instability persists
for a sufficiently long time.

We point out that due to the nature of the approximation used
for the neutrino effects in this study, we can only accurately model
the GW emission in the collapse, bounce, and early postbounce epoch of
the core-collapse supernova scenario, but not much later than
the neutrino burst at shock breakout a few ms after bounce. In future
work we plan to improve upon this and carry out longer-term postbounce
evolutions, where additional GW emission mechanisms may play an
important role~\cite{mueller_04_a,ott_06_a}.

%%%%%%%%%%%%%%%%%%%%%%%%%%%%%%%%%%%%%%%%%%%%%%%%%%%%%%%%%%%%%%%%%%%%%%%%%%%%%%%%%%%
%%%%%%%%%%%%%%%%%%%%%%%%%%%%%%%%%%%%%%%%%%%%%%%%%%%%%%%%%%%%%%%%%%%%%%%%%%%%%%%%%%%
%%%%%%%%%%%%%%%%%%%%%%%%%%%%%%%%%%%%%%%%%%%%%%%%%%%%%%%%%%%%%%%%%%%%%%%%%%%%%%%%%%%

We thank A.~Burrows, E.~M\"uller, S.~Ou, L.~Rezzolla, E.~Seidel,
D.~Shoemaker, N.~Stergioulas, and J.~Tohline for help and stimulating
discussions.  This research was partially supported by the DFG
(SFB/Transregio~7) and by the NCSA under grant No.~AST050022N.

%%%%%%%%%%%%%%%%%%%%%%%%%%%%%%%%%%%%%%%%%%%%%%%%%%%%%%%%%%%%%%%%%%%%%%%%%%%%%%%%%%%
%%%%%%%%%%%%%%%%%%%%%%%%%%%%%%%%%%%%%%%%%%%%%%%%%%%%%%%%%%%%%%%%%%%%%%%%%%%%%%%%%%%
%%%%%%%%%%%%%%%%%%%%%%%%%%%%%%%%%%%%%%%%%%%%%%%%%%%%%%%%%%%%%%%%%%%%%%%%%%%%%%%%%%%

\end{document}